\documentclass[aps,pra,twocolumn,showpacs,superscriptaddress,longbibliography]{revtex4-1}
\usepackage{graphicx} %Include figure files,superscriptaddress
\usepackage{amsmath}
\usepackage{graphicx,epstopdf}
\usepackage[flushleft]{threeparttable}
\usepackage{gensymb}
\newcommand{\be}{\begin{equation}}
\newcommand{\ee}{\end{equation}}
\newcommand{\bea}{\begin{eqnarray}}
\newcommand{\eea}{\end{eqnarray}}
\newcommand{\bse}{\begin{subequations}}
	\newcommand{\ese}{\end{subequations}}

\usepackage{color}
\usepackage[colorlinks,bookmarks=false,citecolor=darkblue,linkcolor=red,urlcolor=blue]{hyperref}

\definecolor{darkred}{rgb}{0.7,0.0,0.0}

\definecolor{darkblue}{rgb}{0,0.02,0.45}

\definecolor{darkgreen}{rgb}{0.02,0.45,0.0}

\definecolor{violet}{rgb}{0.8,0.2,0.6}

\begin{document}

\title{Quasi two-dimensional magnetism in spin-$\frac{1}{2}$ square lattice compound Cu[C$_6$H$_2$(COO)$_4$][H$_3$N-(CH$_2$)$_2$-NH$_3$]$\cdot$3H$_2$O}

\author{S. Guchhait}
\affiliation{School of Physics, Indian Institute of Science Education and Research Thiruvananthapuram-695551, Kerala, India}
\author{S. Baby}
\affiliation{Department of Chemistry, Christian College Chengannur, Alappuzha, Kerala-689122, India}
\author{M. Padmanabhan}
\affiliation{Department of Chemistry, Amrita Vishwa Vidyapeetham, Amritapuri, Kerala-690525, India}
\author{A. Medhi}
\author{R. Nath}
\email{rnath@iisertvm.ac.in}
\affiliation{School of Physics, Indian Institute of Science Education and Research Thiruvananthapuram-695551, Kerala, India}
\date{\today}

\begin{abstract}
We report the crystal growth and structural and magnetic properties of quasi two-dimensional $S=1/2$ quantum magnet Cu[C$_6$H$_2$(COO)$_4$][H$_3$N-(CH$_2$)$_2$-NH$_3$]$\cdot$3H$_2$O. It is found to crystallize in a monoclinic structure with space group $C2/m$. The CuO$_4$ plaquettes are connected into a two-dimensional framework in the $ab$-plane through the anions of [C$_6$H$_2$(COO)$_4$]$^{4-}$ (pyromellitic acid). The [H$_3$N-(CH$_2$)$_2$-NH$_3$]$^{2+}$$\cdot$3H$_2$O groups are located between the layers and provide a weak interlayer connection via hydrogen (H...O) bonds. The temperature dependent magnetic susceptibility is well described by $S=1/2$ frustrated square lattice ($J_1-J_2$) model with nearest-neighbor interaction $J_1/k_{\rm B} \simeq 5.35$~K and next-nearest-neighbor interaction $J_2/k_{\rm B} \simeq -0.01$~K. Even, our analysis using frustrated rectangular lattice ($J_{1a,b}-J_2$) model confirms almost isotropic nearest-neighbour interactions ($J_{\rm 1a}/k_{\rm B} \simeq 5.31$~K and $J_{\rm 1b}/k_{\rm B} \simeq 5.38$~K) in the $ab$-plane and $J_2/k_{\rm B}\simeq-0.24$~K. Further, the isothermal magnetization at $T=1.9$~K is also well described by a non-frustrated square lattice model with $J_1/k_{\rm B} \simeq 5.2$~K. Based on the $J_2/J_1$ ratio, the compound can be placed in the N\'{e}el antiferromagnetic state of the $J_1 - J_2$ phase diagram. No signature of magnetic long-range-order was detected down to 2~K.
\end{abstract}

\maketitle

\section{Introduction}
%The magnetic long-range-ordering (LRO) in two-dimensional (2D) spin systems with continuous symmetry is prevented at a finite temperature.\cite{Mattistheory,Mermin1133} But in real materials exhibit magnetic LRO for the quasi-2D behavior at very low temperature due to the weak inter-plane exchange interaction ($J_{\rm c} \gg J_{\rm \perp}$).
%For a perfect 2D Antiferromagnetic (AF) spin system, susceptibility has a broad maxima at a temperature nearly the average thermodynamic energy scale ($J_{\rm c}$) without any singular cusp associate with magnetic LRO at finite temperature. This properties has many experimental evidence as well as
Quasi-two-dimensional (2D) antiferromagnets are ideal materials to study the interplay between quantum fluctuations and magnetic frustration due to competing interactions. Frustrated square lattice (FSL or $J_1 - J_2$ model) model is the best known example in this category. The Hamiltonian of the isotropic FSL model can be written as
\begin{equation}\label{Heis-Hamil} 
\hat{\mathcal{H}}=J_1\sum_{\langle ij\rangle_1}^{N} S_i\cdot S_j +J_2\sum_{\langle ij\rangle_2}^{N}S_i\cdot S_j,
\end{equation}
where $J_1$ and $J_2$ are the nearest-neighbour (NN) (along the edge) and next-nearest-neighbour (NNN) (along the diagonal) interactions, respectively in a square. The classically possible ground states in this model are determined by the frustration angle $\phi =\text{tan}^{-1}(J_{2}/J_{1})$.\cite{Shannon599} 
There are three possible order states: N\'{e}el antiferromagnetic (NAF, $-0.5\pi \leq \phi \leq 0.15\pi$), columnar antiferromagnetic (CAF, $0.15\pi \leq \phi \leq 0.85\pi$), and ferromagnetic (FM, $0.85\pi \leq \phi\leq-0.5\pi$) states with wave vectors ($Q_x$, $Q_y$) = ($\pi$, $\pi$), [($\pi$, $0$) or ($0$, $\pi$)], and ($0$, $0$), respectively.\cite{Shannon027213,Schmidt214443} The transition regimes NAF/CAF and CAF/FM are known as quantum critical regimes, though the precise boundaries of these regimes are not yet well defined. It is proposed that the ground state in these critical regimes are not exactly quantum spin-liquid but different dimer phases with a singlet gap and gapless nematic phases, respectively.\cite{Sushkov104420,Shannon027213,Singh7278,Capriotti212402,Schmidt125113,Schmidt145211}

%Here, $J_{1}=J_{\rm C}\cos\phi$ (exchange coupling strength along the edges of the square), and $J_{2}= J_{\rm C}\sin\phi$ (exchange coupling strength along the diagonals of the square). The thermodynamic energy scale is fixed by $J_{\rm C}=\sqrt{J_{1}^2+J_{2}^2}$. In these systems, the general behavior of the ordering temperature $T_{\rm N} (\phi,J_{\rm \perp})$ as function of frustration control parameter $\phi =\text{tan}^{-1}(J_{2}/J_{1})$. The inter layer coupling $J_{\rm \perp}(\leq\leq J_{1},J_{2})$ is very weak.
%The ordered moment is determined by the interplay of quantum fluctuation and frustration and may be completely suppressed on approaching small intervals of $\phi$ (0.15$\pi$) around the classical phase boundaries where a quantum spin liquid (QSL) state or more exotic order is expected.

The $J_1-J_2$ phase diagram has been extended further to the spatially anisotropic square lattice or rectangular lattice (known as $J_{1a,b}-J_2$ model).\cite{Schmidt075123} The Hamiltonian for a 2D $S=1/2$ frustrated rectangular lattice (FRL) model can be written as
\begin{equation}\label{Heis-Hamil Rect}
\hat{\mathcal{H}}=J_{1a}\sum_{\langle ij\rangle_{1a}}^{N} S_i\cdot S_j+J_{1b}\sum_{\langle ij\rangle_{1b}}^{N} S_i\cdot S_j +J_2\sum_{\langle ij\rangle_2}^{N}S_i\cdot S_j.
\end{equation}
Here, $J_{\rm 1a}$ and $J_{\rm 1b}$ are the anisotropic exchange couplings along the edges of the square and the coupling along the diagonals ($J_2$) remains same. The classically predicted phase diagram becomes a function of frustration angle $\phi=\text{tan}^{-1}\left(J_{2}/\sqrt{\frac{(J_{1a}^{2} + J_{1b}^{2})}{2}}\right)$ and anisotropy parameter $\theta = \text{tan}$$^{-1}(J_{1b}/J_{1a})$.
%$J_{1a}=\sqrt{2}J_{\rm C}\cos\phi\cos\theta$, $J_{1b}=\sqrt{2}J_{\rm C}\cos\phi\sin\theta$, and $J_{\rm C}=\sqrt{\frac{(J_{1a}^{2}+J_{1b}^{2})}{2}+J_{2}^{2}}$.
The introduction of a rectangular distortion does not significantly change the phase diagram. The predicted phases are FM, NAF, and columnar antiferromagnets [CAF$_a$, ($Q_x$, $Q_y$) = ($\pi$, 0) and CAF$_b$, ($Q_x$, $Q_y$) = (0, $\pi$)]. The only difference is that the CAF phases are degenerate for the isotropic model $(J_{1a}=J_{1b})$ with $\theta=\frac{\pi}{4}$ or $\theta=\frac{-3\pi}{4}$. Further, the $J_{1a,b}-J_2$ model predicts that the CAF phase is stable for all values of $\phi$, especially in the spin nematic phase regime of the isotropic $J_{1} -J_{2}$ model.\cite{Schmidt075123}

%\begin{figure}
%	\includegraphics[]{Fig_1} 
%	\caption{ Classical phase diagram of $J_{1}-J_{2}$ model for spin-1/2 systems. Position of BaCdVO(PO$_4$)$_2$ is expected only presence of magnetic field 3.8~T$<H\leq4.5$~T. VO(HCOO)$_2$·(H$_2$O) and C$_{12}$H$_{18}$CuN$_{2}$O$_{11}$ have two positions in NAF region. The value of $J_1/J_2$ ratio taken from: Zn$_2$VO(PO$_4$)$_2$,~\cite{Yogi024413} Sr$_2$CuTeO$_6$,~\cite{Mustonen1085}, VOMoO$_4$~\cite{Bombardi220406}, PbVO$_3$~\cite{Tsirlin092402}, Sr$_2$CuWO$_6$,~\cite{Mustonen1085} Sr$_2$CuW$_0.5$Te$_0.5$O$_6$,~\cite{Mustonen1085} RbMoOPO$_4$Cl~\cite{Ishikawa064408}, KMoOPO$_4$Cl~\cite{Ishikawa064408}, Na$_1.5$VO(PO$_4$)F~\cite{Tsirlin014429}, (CuBr)Sr$_2$Nb$_3$O$_10$~\cite{Tsujimoto63711}, (CuBr)LaNb$_2$O$_7$.~\cite{Tsujimoto63711} }
%	\label{Fig_1}
%\end{figure}

The $S=1/2$ FSL model has been realized in the class of 
layered V$^{4+}$ based inorganic compounds $AA^{\prime}$(VO)(PO$_4$)$_2$ ($AA^{\prime}$ = Zn$_2$, Pb$_2$, SrZn, PbZn, BaZn, and BaCd) and Li$_2$VO(Si,Ge)O$_4$.\cite{Nath064422,Nath214430,Yogi024413,Tsirlin174424,Carretta224432,Bossoni014412,Nath214430,Roy012048,Nath064422,Tsirlin174424,Rosner014416,Bettler184437,Tsirlin014429,Tsirlin132407} Among these compounds, BaCdVO(PO$_4$)$_2$ is the one located very close to the nematic phase regime in the $J_1 - J_2$ phase diagram and is being extensively studied. Some of the recent studies have reported the signature of spin nematic phase in BaCdVO(PO$_4$)$_2$.\cite{Povarov024413,Skoulatos014405,Bhartiya033078}
%The investigation on single crystal by Povarov~\textit{et.al.} quantified “dimensionality reduction” effect serving as the indicator of strong frustration close to the saturation field region at low temperature. They also found the spin-nematic ground state in presence of applied field.\cite{Povarov024413} The field dependent neutron-diffraction and ac magnetic susceptibility measurements of this system realized the presence of bond spin nematic (BSN) phase in BaCdVO(PO$_4$)$_2$ in the field range 3.8~T$<H\leq 4.5$~T.\cite{Skoulatos014405,Bhartiya033078}
A few metal-organic compounds based on V$^{4+}$ and Cu$^{2+}$ have also been studied in light of the 2D spin-$1/2$ Heisenberg model.\cite{Guchhait104409,Woodward144412,Ronnow037202,Nath054409}
A series of Cu based quasi-2D organometallic magnets where Cu$^{2+}$ ions are bridged by pyrazine molecules are [Cu(HF$_2$)(pyz)$_2$]$X$ ($X$ = BF$_4^-$, ClO$_4^-$, PF$_6^-$, SbF$_6^-$, and AsF$_6^-$)\cite{Goddard083025} and [Cu(pyz)$_2$]$X_2$ ($X$= ClO$_4^-$ and BF$_4^-$).\cite{Lancaster094421,Woodward4256,Tsyrulin134409} These compounds are having square lattice network with negligible NNN exchange coupling ($J_{2}$). Another family of Cu based organo-metallic square lattice compounds are $A_2$Cu$X_4$ ($A$ = 5CAP and 5MAP, $X$ = Br and Cl) without frustration.\cite{Woodward144412}
%C$_{12}$H$_{18}$CuN$_{2}$O$_{11}$ is formed by [Cu(C$_6$H$_2$(COO)$_4$)]$^2-$ layers (rectangular arrays of Cu$^{2+}$ ions) connected by \{[$^{+}$H$_3$N-(CH$_2$)$_2$-NH$_3^{+}$].2H$_2$O\} (organic cationic group) ligands, which is similar to the structure of Cu[C$_6$H$_2$(COO)$_4$][C$_2$H$_5$NH$_3$]$_2$ (Cu(PM)(EA)$_2$).
Recently, we have reported that Cu[C$_6$H$_2$(COO)$_4$][C$_2$H$_5$NH$_3$]$_2$ is a quasi-2D  spatially anisotropic non-frustrated spin-$1/2$ square lattice with exchange couplings $J_{1a}/k_{\rm B}=5.6$~K and $J_{1c}/k_{\rm B}=8.0$~K along $a$- and $c$-directions, respectively.\cite{Nath054409}
 
In this work, we report the synthesis and magnetic properties of a new organic spin-1/2 quantum magnet Cu[C$_6$H$_2$(COO)$_4$][H$_3$N-(CH$_2$)$_2$-NH$_3$]$\cdot$3H$_2$O (or C$_{12}$H$_{18}$CuN$_{2}$O$_{11}$). The magnetization data analysis confirms the non-frustrated quasi-2D nature with a weak anisotropy in the in-plane couplings. It does not show the onset of magnetic long-range-ordering (LRO) down to 2~K, reflecting weak inter-plane coupling and hence perfect two-dimensionality. 

\begin{figure}
	\includegraphics[scale=0.42]{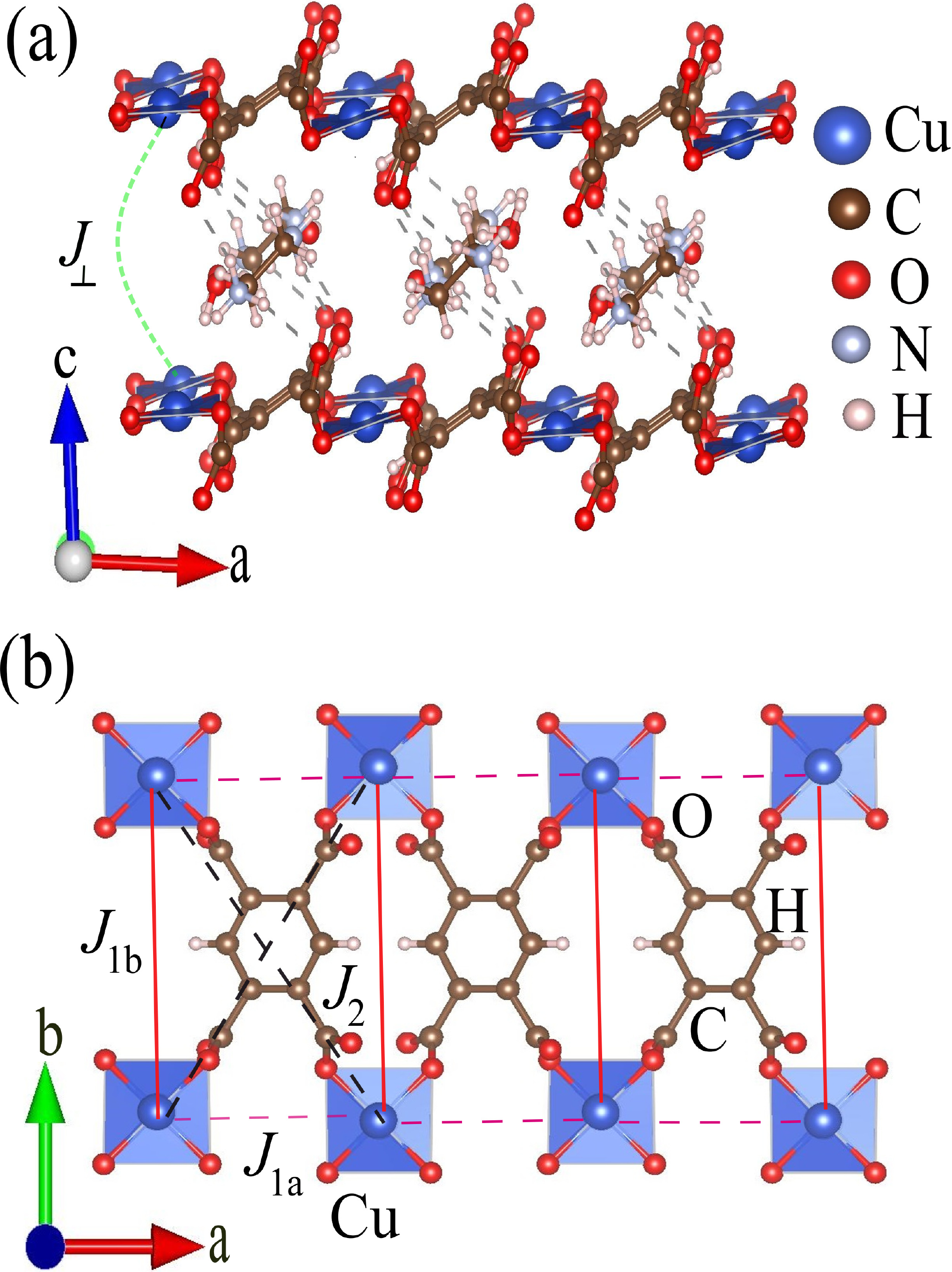} 
	\caption{(a) Three dimensional view of the C$_{12}$H$_{18}$CuN$_{2}$O$_{11}$ structure featuring negatively charged \{Cu[C$_6 $H$_2 $(COO)$_4$]\}$^{2-}$ layers connected by the [H$_3$N-(CH$_2$)$_2$-NH$_3$]$^{2+}$$\cdot$3H$_2$O groups through hydrogen bond. $J_{\rm \perp}$ is the exchange coupling between two layer. (b) A section of the \{Cu[C$_6 $H$_2 $(COO)$_4$]\}$^{2-}$ layer in $ab$-plane showing the exchange couplings forming a rectangular spin lattice of Cu$^{2+}$ ions. The exchange couplings $J_{\rm 1a}$ and $J_{\rm 1b}$ are along the edges of the rectangle and $J_{\rm 2}$ is along the diagonal.}
    \label{Fig1}
\end{figure}

\section{Techniques}
Single crystals of the Cu(II)-based metal organic hybrid compound C$_{12}$H$_{18}$CuN$_{2}$O$_{11}$ were synthesised by using 1,2,4,5-benzenetetracarboxylic acid (H$_4$BTC). Since the compound contains four carboxylic acid groups, we were initially getting mixture of products from which isolation of pure phase of the material was difficult. After repeated trials and by varying the reaction conditions the phase-pure form of the compound was obtained by adopting the following procedure. Copper acetate monohydrate (5 mmol, 1.00 g), ethylene diamine (5 mmol, 0.35 mL), H$_4$BTC (5 mmol, 1.27 g) were reacted in 30 mL DMF-water mixture (taken in 1:1 volume ratio). The initial blue product formed was filtered out. The clear and pale blue filtrate obtained was kept for slow evaporation for 8 days at room temperature. Light bluish needle type crystals of the target compound in phase-pure form were separated and dried in air. The yield was 45\% (based on Cu).

%\begin{figure}
%\includegraphics[scale=0.5]{Fig2.eps}
%\caption{Powder XRD patterns (open circles) at room temperature (300~K) and 15~K for C$_{12}$H$_{18}$CuN$_{2}$O$_{11}. The solid line is the Rietveld fit, the vertical bars represent the expected Bragg peak positions, and the lower solid line corresponds to the difference between the observed and calculated intensities.}
%\label{Fig_2}
%\end{figure}
Single crystal x-ray diffraction (XRD) was performed on a good-quality single crystal at room temperature using a Bruker KAPPA APEX-II CCD diffractometer equipped with graphite monochromated Mo $K_{\alpha1}$ radiation ($\lambda=0.71073$~\AA). The data were collected using  APEX3 software and reduced with SAINT/XPREP.\cite{Bruker2016apex3} An empirical absorption correction was done using the SADABS program.\cite{Sheldrick1994} The structure was solved with direct methods using SHELXT-2018/2\cite{Sheldrick2015shelxt} and refined by the full matrix least squares on $F^{2}$ using SHELXL-2018/3, respectively.\cite{Sheldrick2018shelxl} All the hydrogen atoms were placed geometrically and held in the riding mode for the final refinements. The final refinements included atomic positions for all the atoms, anisotropic thermal parameters for all the nonhydrogen atoms, and isotropic thermal parameters for the hydrogen atoms. The crystal data and details of the structure refinement parameters are listed in Table~\ref{Structure1}. 
%To further confirm the phase purity, a large number of single crystals were crushed into powder, and powder XRD was performed at room temperature using a PANalytical (Cu $K_\alpha$ radiation, $\lambda_{\rm ave}=1.54060$~\AA) diffractometer. Rietveld refinement of the powder XRD pattern was performed using FULLPROF package, taking the initial structural parameters from the single crystal data (as in Table~\ref{Structure1}).\cite{Carvajal55} The obtained best fit parameters are $a \simeq xx$~\AA, $b \simeq xx$~\AA, $c \simeq xx$~\AA, $\beta = xx^{\circ}$ and $V_{\rm cell} \simeq xx$~\AA$^{3}$, and the goodness-of-fit $\chi^{2}\simeq xx$. These lattice parameters are consistent with the single crystal data.

As the size of the crystals was too small, it was not possible to do the magnetic measurements on the individual crystals and hence powder sample was used for this purpose. The temperature ($T$) dependent magnetic susceptibility [$\chi(T)$] in four different magnetic fields ($\mu _{\rm 0}H= 0.5$, 1, 3, and 5~T) was measured in the temperature range $2 \leq T\leq 300$~K using the vibrating sample magnetometer (VSM) attachment to the Physical Property Measurement System (PPMS, Quantum Design). A magnetic isotherm (magnetization $M$ vs field $H$) was measured by varying the magnetic field from 0 to 14 T at $T=1.9$~K.

The Quantum Monte Carlo (QMC) simulation for magnetization was performed assuming the Heisenberg model on a nonfrustrated square lattice with an isotropic exchange coupling. We used the Hamiltonian in the presence of a magnetic field $\hat{\mathcal{H}}=J\sum_{\langle i,j\rangle}S_i\cdot S_j-H\sum_{i}S^{z}_{i}$,
where $J$ represents the exchange coupling strength between spins at the $i^{th}$ and $j^{th}$ sites and $H$ is the external magnetic field. 
We used the directed loop QMC algorithm in the stochastic series expansion representation\cite{SandvikR14157,Alet036706} implemented in the ALPS software package.\cite{ALPS}
The lattice size was taken to be $20\times 20$ (400 sites) and measurements were done from a simulation of about $10^5$ sweeps including about 5000 thermalization sweeps.

\section{Results}
\subsection{Crystal Structure}
\begin{table}[ptb]
	\caption{Crystal structure data for C$_{12}$H$_{18}$CuN$_{2}$O$_{11}$at room temperature.}
	\label{Structure1}
			\begin{tabular}{ccc}
			\hline \hline
			Empirical formula & C$_{12}$H$_{18}$CuN$_{2}$O$_{11}$ & \\
			Formula weight ($M_r$) & 429.8 & \\
			Temperature & 296(2) K\\
			Crystal system & Monoclinic & \\
			Space group & $C2/m$ & \\
			Lattice parameters & $a=11.4258(3)$~\AA, & \\
			& $b=18.4562(5)$~\AA, & \\
			& $c=7.4747(2)$~\AA,& \\
			& $\beta=95.079(2)^{\circ}$ & \\	
			Unit cell volume ($V_{\rm cell}$) & 1570.05(7)~\AA$^3$ & \\       
			Z & 4\\ 
			Radiation type & Mo$K_{\alpha1}$\\    	
			Wavelength ($\lambda$) & 0.71073~\AA \\
			Diffractometer & Bruker KAPPA APEX-II CCD \\	
			Crystal size & $0.2 \times 0.15 \times 0.1$ mm$^3$\\
			2$\Theta$ range for data collection & 4.2$^{\circ}$ to 50$^{\circ}$\\
			Index ranges & $-13\leq h\leq 13$, & \\
			& $-21\leq k\leq 21$, & \\
			& $-8\leq l\leq 8$ & \\
			Absorption coefficient ($\mu$) & 1.459 mm$^{-1}$\\
			$F$(000) & 884\\
			Reflections collected & 6671\\
			Independent reflections & 1429 [$R_{\rm int} = 0.0183$] \\
			Data/restraints/parameters & 1429/3/128\\
			Goodness-of-fit on $F^{2}$ & 1.104\\
			Final $R$ indexes, $I\geq 2\sigma(I)$ & $R_{1}=0.0272$, $\omega R_{2} = 0.0709$& \\
			Final $R$ indexes, all data & $R_{1}=0.0293$, $\omega R_{2}=0.0723$ & \\
			Largest difference peak/hole & 1.014 / -0.487 e.\AA$^{-3}$ \\
			Calculated crystal density $\rho_{\rm cal}$ & 1.818 mg/mm$^3$\\
			\hline \hline 
		\end{tabular}
\end{table}
C$_{12}$H$_{18}$CuN$_{2}$O$_{11}$ stabilizes in a monoclinic crystal structure with space group $C2/m$. The lattice parameters, atomic positions, and main bond distances along with their angles at room temperature are tabulated in Tables~\ref{Structure1}, \ref{atmic position}, and \ref{bond length}, respectively. The crystal structure is shown in Fig.~\ref{Fig1}. Each Cu atom is bonded with four O atoms forming a CuO$_4$ square. As the Cu-O distances are unequal, CuO$_4$ is slightly distorted.
The CuO$_4$ plaquettes are connected via [C$_6$H$_2$(COO)$_4$]$^{4-}$ building rectangular layers in the $ab$-plane [Fig.~\ref{Fig1}(b)]. The distance between NN Cu$^{2+}$ ions along the smaller edge (along $a$-axis) of a rectangle is $\sim 5.7176$~\AA while along the longer edge (along $b$-axis) these distances are unequal ($\sim 8.9963$~\AA ~and $\sim 9.4599$~\AA). Hence, the rectangular lattice is expected to be anisotropic or to form a trapezoid. The corresponding exchange couplings are marked as $J_{\rm 1a}$ and $J_{\rm 1b}$ along the $a$- and $b$-axes, respectively as shown in Fig.~\ref{Fig1}(b).
The NNN distances between Cu$^{2+}$ ions along diagonals of the rectangle is $\sim 10.8533$~\AA with exchange coupling $J_{\rm 2}$. 
Further, the distance between two Cu$^{2+}$ ions in two adjacent layers along the crystallography $c$-axis is $\sim 7.4747$~\AA. The  [H$_3$N-(CH$_2$)$_2$-NH$_3$]$^{2+}$$\cdot$3H$_2$O groups lie sandwiched between the layers and are connecting the Cu$^{2+}$ ions from the adjacent layers via weak hydrogen bonds [see Fig.~\ref{Fig1}(a)]. Thus, because of the large spacial distance and weak hydrogen bonding, the inter-layer interaction ($J_{\perp}$) is expected to be very weak.
\begin{table}
	\setlength{\tabcolsep}{0.2cm}
	\caption{The atomic coordinates ($x,y,z$) for C$_{12}$H$_{18}$CuN$_{2}$O$_{11}$. $U_{\rm iso}$ is the isotropic atomic displacement parameters which is defined as one-third of the trace of the orthogonal $U_{\rm ij}$ tensor. The errors are from the least-square structure refinement. The positions of hydrogen atoms are fixed.}
	\label{atmic position}
	\begin{tabular}{ccccccc}
		\hline \hline
		Atomic sites & $x$ & $y$ & $z$ & $U_{\rm iso}$(\AA$^{2}$) & \\\hline
		   Cu(1) & 0.5000 & 0.2563(1) & 0.5000 & 0.014(1)\\
		   C(1) & 0.6438(2) & 0.3652(1) & 0.4217(3) & 0.017(1)\\            
		   C(2) & 0.7105(2) & 0.4346(1) & 0.4625(3) & 0.016(1)\\            
		   C(3) & 0.8186(2) & 0.4346(1) & 0.5676(3) & 0.015(1)\\          
		   C(4) & 0.8826(2) & 0.3647(1) & 0.6121(3) & 0.017(1)\\           
	       C(5) & 0.6582(3) & 0.5000 & 0.4095(5) & 0.018(1)\\          
		   C(6) & 0.8710(3) & 0.5000 & 0.6194(5) & 0.017(1)\\ 
		   C(7) & 0.8001(3) & 0.2579(2) & 1.0704(4) & 0.028(1) \\            
		   N(1) & 0.8215(2) & 0.3372(1) & 1.0812(3) & 0.027(1)\\                
		   O(1$^\prime$) & 0.6135(3) & 0.4228(2) & 0.8964(3) & 0.053(1)\\
		   O(2$^\prime$) & 1.1385(14) & 0.5000 & 0.9284(18) & 0.276(7)\\      
		   O(1) & 0.6159(1) & 0.3316(1) & 0.5616(2) & 0.019(1)\\       
		   O(2) & 0.6151(2) & 0.3465(1) & 0.2655(2) & 0.029(1)\\        
		   O(3) & 0.8804(1) & 0.3193(1) & 0.4837(2) & 0.021(1)\\           
		   O(4) & 0.9361(2) & 0.3563(1) & 0.7637(2) & 0.026(1)\\             
		   H(5) & 0.5875 & 0.5000 & 0.3377 & 0.021\\  
		   H(6) & 0.9423 & 0.5000 & 0.6899 & 0.021\\           
		   H(1A) & 0.8805 & 0.3462 & 1.1641 & 0.04\\	
		   H(1B) & 0.8399 & 0.3534 & 0.9751 & 0.04\\	
		   H(1C) & 0.7569 & 0.3595 & 1.1109 & 0.04\\	
		   H(7A) & 0.7809 & 0.2399 & 1.1862 & 0.034\\
		   H(7B) & 0.8706 & 0.2333 & 1.0394 & 0.034\\	
		   H(1A$^\prime$) & 0.581(3) & 0.407(4) & 0.784(4) & 0.14(3)\\
		   H(1B$^\prime$) & 0.6826 & 0.4369 & 0.8998 & 0.21(4)\\	
		\hline\hline
	\end{tabular}
\end{table}
\begin{table}
	\setlength{\tabcolsep}{0.00001cm}
	\caption{Some selected bond lengths and bond angles for C$_{12}$H$_{18}$CuN$_{2}$O$_{11}$.}
	\label{bond length}
	\begin{tabular}{ccccccc}
		\hline \hline
			& Bond length & &Bond length \\
			& (\AA) & & (\AA) \\\hline
		C(1)-O(2) & 1.234(3) & C(4)-O(4) & 1.249(3)\\
		C(1)-O(1) & 1.280(3) & C(4)-O(3) & 1.273(3)\\
		C(1)-C(2) & 1.508(3) & N(1)-C(7) & 1.485(4)\\
		C(2)-C(5) & 1.389(3) & C(7)-C(7)$^1$ & 1.513(5)\\
		C(2)-C(3) & 1.404(3) & O(1)-Cu(1) & 1.9464(16)\\
		C(3)-C(6) & 1.388(3) & O(3)-Cu(1)$^2$ & 1.9490(16)\\
		C(3)-C(4) & 1.505(3)\\
		& Bond angles & & Bond angles \\
		& ($^{\circ}$) & & ($^{\circ}$) \\
		O(2)-C(1)-O(1) & 124.9(2) & C(2)$^3$-C(5)-C(2) & 120.7(3)\\
		O(2)-C(1)-C(2) & 121.1(2) & C(3)$^3$-C(6)-C(3) & 121.0(3)\\
		O(1)-C(1)-C(2) & 113.9(2) & N(1)-C(7)-C(7)$^1$ & 109.8(3)\\
		C(5)-C(2)-C(3) & 119.6(2) & C(1)-O(1)-Cu(1) & 111.50(15)\\
		C(5)-C(2)-C(1) & 118.9(2) & C(4)-O(3)-Cu(1)$^2$ & 117.21(15)\\
		C(3)-C(2)-C(1) & 121.1(2) & O(1)-Cu(1)-O(1)$^4$ & 88.92(10)\\
		C(6)-C(3)-C(2) & 119.5(2) & O(1)-Cu(1)-O(3)$^5$ & 169.86(7)\\
		C(6)-C(3)-C(4) & 119.6(2) & O(1)$^4$-Cu(1)-O(3)$^5$ & 92.14(7)\\
		C(2)-C(3)-C(4) & 120.7(2) & O(1)-Cu(1)-O(3)$^2$ & 92.14(7)\\
		O(4)-C(4)-O(3) & 125.2(2) & O(1)$^4$-Cu(1)-O(3)$^2$ & 169.86(7)\\
		O(4)-C(4)-C(3) & 119.8(2) & O(3)$^5$-Cu(1)-O(3)$^2$ & 88.59(10)\\
		O(3)-C(4)-C(3) & 114.9(2)\\		
		\hline\hline 
	\end{tabular}
	\begin{tablenotes}
		\item Symmetry transformations used to generate equivalent atoms of table~\ref{bond length}:\\ 
		$^1$-x+3/2,-y+1/2,-z+2 $^2$-x+3/2,-y+1/2,-z+1 $^3$x,-y+1,z \\
		$^4$-x+1,y,-z+1 $^5$x-1/2,-y+1/2,z. 
	\end{tablenotes}
\end{table}

\subsection{Magnetic Susceptibility} 
\begin{figure}
	\includegraphics[scale=1]{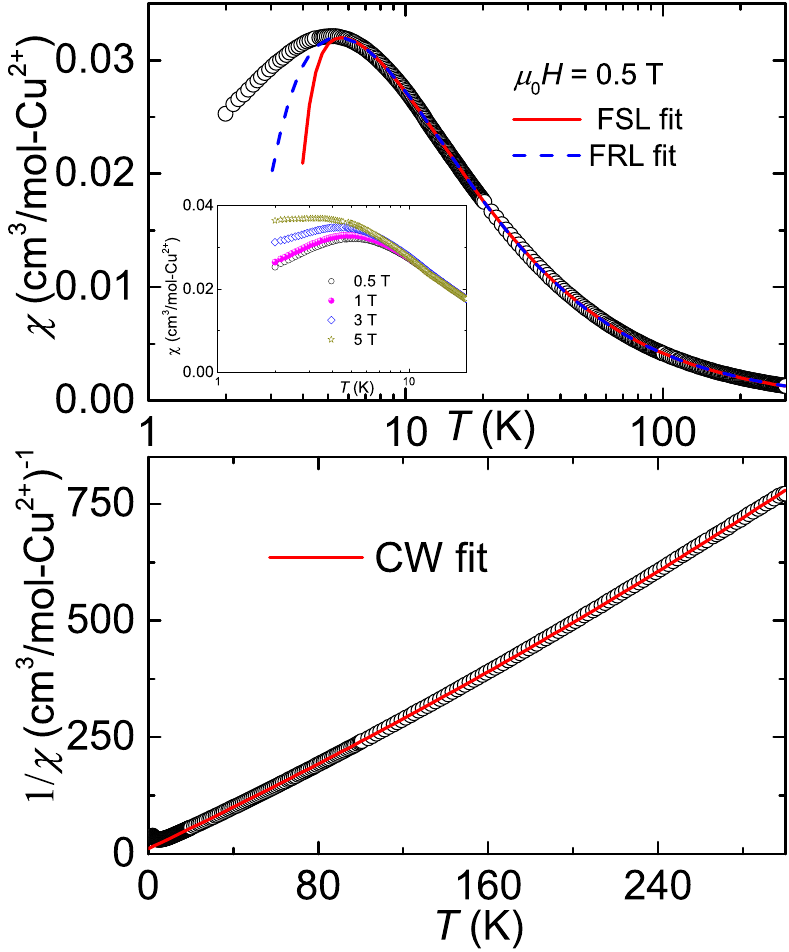}
	\caption{Upper panel: $\chi(T)$ vs $T$ in an applied field of $\mu_{\rm 0}H = 0.5$~T. The solid and dashed lines are the best fits of the data using HTSE of frustrated square lattice and frustrated rectangular lattice models [Eq.~\eqref{chi_twoD}], respectively. Inset: The low temperature $\chi(T)$ measured in different fields. Lower panel: Inverse magnetic susceptibility ($1/\chi$) vs $T$ and the solid line is the Curie-Weiss fit.}
	\label{Fig2}
\end{figure}
Magnetic susceptibility ($\chi= M/H$) as a function of temperature ($T$) measured in an applied field of $\mu_{0}H = 0.5$~T is shown in the upper panel of Fig.~\ref{Fig2}. In the high temperature region, $\chi(T)$ increases systematically with lowering temperature, typically expected in the paramagnetic state. It then passes through a broad maximum at around $T_{\chi}^{\rm max} \simeq 5.13$~K mimicking the short-range AF ordering in the system. This is a clear evidence of quasi-2D nature of the compound. No signature of magnetic LRO was observed down to 2~K. As shown in the inset of the upper panel of Fig.~\ref{Fig2}, the broad maximum shifts towards lower temperatures with increasing magnetic field. This behavior is quite similar to that observed in other low-dimensional antiferromagnets.\cite{Nath054409,Nath064422}

$\chi(T)$ in the high temperature region can be fitted by
\begin{equation}\label{CW}
\chi(T) = \chi_0 + \frac{C}{T - \theta_{\rm CW}},
\end{equation}
where, $\chi_0$ is the temperature-independent susceptibility consisting of core diamagnetic susceptibility ($\chi_{\rm dia}$) of the core electron shells of the atoms and Van-Vleck paramagnetic susceptibility ($\chi_{\rm vv}$) of the open shells of the Cu$^{2+}$ ions in the sample. The second term is the Curie-Weiss (CW) law where $C$ is Curie constant and $\theta_{\rm CW}$ is Curie-Weiss temperature. Our experimental $\chi(T)$ data in the temperature range $T \geq 18$~K were fitted well by Eq.~\eqref{CW} yielding $\chi_0 \simeq -2.26\times 10^{-4}$~cm$^3$/mol-Cu$^{2+}$, $C\simeq0.46$~cm$^3$.K/mol-Cu$^{2+}$, and $\theta_{\rm CW}\simeq-5.17$~K. The negative Curie-Weiss temperature indicates predominance of AF exchange interactions between the Cu$^{2+}$ ions in the compound. From the value of $C$, the effective magnetic moment $\mu_{\rm eff} =(3k_{\rm B}C/N_{\rm A}\mu_{\rm B}^{2})^{\frac{1}{2}}$, (where $k_{\rm B}$ is the Boltzmann constant, $N_{\rm A}$ is the Avogadro’s number, and $\mu_{\rm B}$ is the Bohr magneton) is estimated to be $\mu_{\rm eff} \simeq 1.91$~$\mu_{\rm B}/$Cu$^{2+}$. This value of $\mu_{\rm eff}$ [$= g\sqrt{S(S+1)}\mu_{\rm B}$] corresponds to a Landé $g$-factor of $g \simeq 2.21$ which is slightly larger than the ideal value ($g = 2$), expected for spin-$1/2$. A slightly larger value of $g$ is typically found for Cu$^{2+}$ based compounds from ESR experiments.\cite{Nath014407,Janson094435,Arango134430}

To understand the geometry of the spin lattice, $\chi(T)$ in the high temperature regime was fitted by the sum of a temperature independent term ($\chi_0$) and a temperature dependent term
\begin{equation}
\label{chi_twoD}
\chi(T) = \chi_0 + \chi_{\rm spin}(T).
\end{equation}
Here, $\chi_{\rm spin}(T)$ is the high-temperature series expansion (HTSE) of spin susceptibility for the spin-$1/2$ FSL model ($J_1-J_2$ model).\cite{Rosner014416,Schmidt104443} The expression is given by
\begin{eqnarray}
\label{chi_spin}
\chi_{\rm spin}(T)=\frac{N_{\rm A}g^{2}\mu_{\rm B}^{2}}{k_{\rm B}T}\sum_{n}\left(\frac{J_{1}}{k_{\rm B}T}\right)^{n}\sum_{m}c_{m,n}\left(\frac{J_{\rm 2}}{J_{\rm 1}}\right)^{m}.
\end{eqnarray}
The values of the coefficients, $c_{m,n}$ are tabulated in Ref.~\cite{Rosner014416}.
The best fit of the $\chi(T)$ data (upper panel of Fig.~\ref{Fig2}) by Eq.~\eqref{chi_twoD} in the temperature range $T > 5.4$~K resulted two different solutions: Solution I: $\chi_0\simeq-2.65\times10^{-4}$~cm$^3$/mol-Cu$^{2+}$, $J_1/k_{\rm B}\simeq5.35$~K, $J_2/k_{\rm B}\simeq-0.01$~K, and $g\simeq2.23$ and Solution II: $\chi_0 \simeq-2.68\times10^{-4}$~cm$^3$/mol-Cu$^{2+}$, $J_1/k_{\rm B}\simeq5.35$~K, $J_2/k_{\rm B}\simeq0.01$~K, and $g\simeq2.23$. As discussed later, the solution I appears to be the correct solution. In both cases, the value of $J_2$ is negligibly small and hence can be ignored. Nevertheless, for both the solutions the compound can be placed in the NAF regime of the $J_1 - J_2$ phase diagram.

As discussed earlier, the Cu$^{2+}$ ions form a slightly distorted square lattice. In an attempt to test the spin-lattice, $\chi(T)$ data were fitted by the FRL model (see Fig.~\ref{Fig2}). The fit was done using Eq.~\eqref{chi_twoD} where $\chi_{\rm spin}$ is taken as HTSE for the anisotropic FSL/FRL model given in Ref.~\cite{Schmidt104443}. Our fit in the temperature range $T > 5.4$~K results $\chi_0\simeq-2.3\times 10^{-4}$~cm$^3$/mol-Cu$^{2+}$, $g\simeq2.22$, $J_{\rm 1a}/k_{\rm B}\simeq5.31$~K, $J_{\rm 1b}/k_{\rm B}\simeq5.38$~K, and $J_{\rm 2}/k_{\rm B}\simeq-0.24$~K. As $J_{\rm 1a}/k_{\rm B}$ and $J_{\rm 1b}/k_{\rm B}$ are having almost equal magnitude, the spin-lattice can essentially be treated as a weakly anisotropic square lattice.

\subsection{Magnetic Isotherm}
\begin{figure}
	\includegraphics {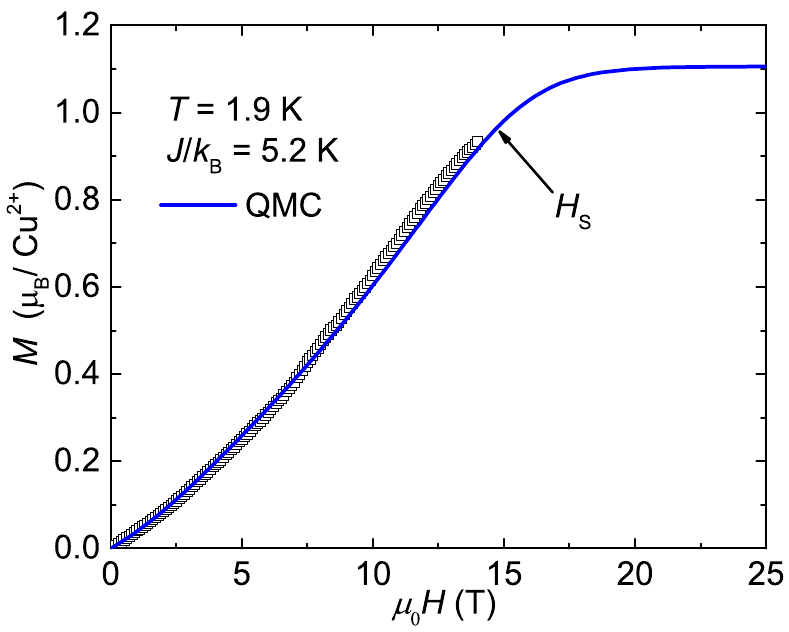}
	\caption {Magnetization $(M)$ as a function of magnetic field $H$ at $T=1.9$~K measured up to 14~T. The solid line is the QMC simulation, assuming a uniform nonfrustrated square lattice model with $J/k_{\rm B} = 5.2$~K.}
	\label{Fig3}
\end{figure}
Magnetization ($M$) as a function of applied field ($H$) measured at $T=1.9$~K is shown in Fig.~\ref{Fig3}. $M$ varies almost linearly with $H$ with a small curvature and at $\mu_{0}H = 14$~T it is still below the saturation field. According to theoretical calculation by Schmidt~\textit{et al.}\cite{Schmidt125113}, the saturation field of a FSL model can be expressed as
\begin{equation}
\begin{split}
\mu_{0}H_{\rm S}=\frac{J_{c}k_{\rm B}zS}{g\mu _{\rm B}} & \left\{\left[1-\frac{1}{2}(\cos
Q_{x}+\cos Q_{y})\right] \cos \phi\right. \\
&\quad \left. {}+(1-\cos Q_{x}\cos Q_{y})\sin \phi\vphantom{\frac12}\right\},
\end{split}
\end{equation}
where $z=4$ is the magnetic coordination number, $S=1/2$, and ($Q_{x}$, $Q_{y}$)
are the wave vectors which are different for different ordered states. Putting ($Q_{x}$, $Q_{y}$) = ($\pi$, $\pi$), the saturation field for the NAF phase will have the form $\mu_{\rm 0}H_{\rm S}= 4J_{\rm 1}k_{\rm B}/(g\mu _{\rm B})$, which is independent of $J_{\rm 2}$. Using $J_{\rm 1}/k_{\rm B}\simeq5.35$~K and $g=2.23$ in this formula, the value of saturation field is calculated to be $\mu_{0}H_{\rm S}^{\rm sq}\simeq 14.3$~T. Even putting the values of $J_{\rm 1a}$  and $J_{\rm 1b}$ in a spin-$1/2$ FRL model, the saturation field is calculated to be $\mu_{0}H_{\rm S}^{\rm rect}= 2(J_{\rm 1a}+J_{\rm 1b})k_{\rm B}/(g\mu _{\rm B})\simeq 14.3$~T.\cite{Nath054409} 

In order to further understand the nature of spin lattice, QMC simulation is done taking $J/k_{\rm B}=5.2$~K in a non-frustrated square lattice model. As shown in Fig.~\ref{Fig3}, the QMC simulated data reproduce the shape of our experimental curve perfectly reflecting the non-frustrated square lattice nature of the spin-lattice. The simulated curve changes the slope at around $\mu_{0}H\simeq15$~T, which is very close to the saturation field expected for the compound. It reaches a saturation magnetization of $M_{\rm S} \simeq 1.1\mu_{\rm B}$/Cu$^{2+}$ for $\mu_{0}H > 15$~T which is consistent with the expected value of $M_{\rm S} = gS\mu_{\rm B}\simeq1.1\mu_{\rm B}/$Cu$^{2+}$ for $S = 1/2$ and $g=2.23$.

\section{Discussion and Summary}
According to mean field approximation, for the FSL model, one can write $\theta_{\rm CW} = \frac {zS(S+1)}{3k_{\rm B}}(J_1+J_2)$.\cite{Domb} Taking $S = 1/2$, $z=4$, $J_1/k_{\rm B} \simeq 5.35$~K, and $J_2/k_{\rm B} \simeq -0.01$~K, we got $\theta_{\rm CW}\simeq 5.34$~K which is very close to the CW temperature obtained from the $1/\chi$ analysis.
Using the values of $J_1$ and $J_2$, the frustration control parameter is calculated to be $\phi =- 0.1^{\circ}$ ($\sim -0.0006\pi$), which places the compound in the NAF ordered state of the $J_1-J_2$ phase diagram.\cite{Nath064422} Similarly, for a FRL model one can write $\theta_{\rm CW}=(\frac{J_{\rm 1a}+J_{\rm 1b}}{2}+J_{\rm 2})/k_{\rm B}$. Taking $J_{\rm 1a}/k_{\rm B} \simeq 5.31$~K, $J_{\rm 1b}/k_{\rm B} \simeq 5.38$~K, and $J_{2}/k_{\rm B} \simeq -0.24$~K we got $\theta_{\rm CW}\simeq 5.11$~K which is even closer to the CW temperature obtained from the $1/\chi$ analysis. The anisotropic angle and frustration angle are estimated to be $\theta\simeq 0.252\pi$ and $\phi\simeq-0.014\pi$, respectively in the NAF regime of the $J_{1a,b}-J_2$ phase diagram.\cite{Schmidt075123}

Usually, in a frustrated magnet, the extent of frustration can be quantified by the frustration parameter $f=\frac{|\theta_{\rm CW}|}{T_{\rm N}}$. C$_{12}$H$_{18}$CuN$_{2}$O$_{11}$ has no magnetic LRO down to 2~K which makes this system a good example of a quasi-2D AF system. The lower limit of the frustration parameter of this compound is estimated to be $f > \frac{5.17}{2}\simeq2.6$, taking the upper limit of $T_{\rm N} = 2$~K. Here, $|\theta_{\rm CW}|>T_{\rm N}$ implies that the magnetic LRO ($T_{\rm N}$) is prevented by quantum fluctuations due to low dimensionality of the spin-lattice and the role of frustration has negligible effect.
Further, assuming that $T_{\rm N} < 2$~K and using the appropriate exchange couplings, the upper limit of the inter-layer coupling is estimated to be negligibly small compared to the intra-layer coupling.\cite{Majlis7872,Schmidt214443}
Thus, this compound is another example of a quasi-2D nonfrustrated system with $J_1/T_{\rm N} > 2.67$, similar to the compounds tabulated in Ref.~\cite{Guchhait104409}.

\begin{figure}
	\includegraphics[scale=.8]{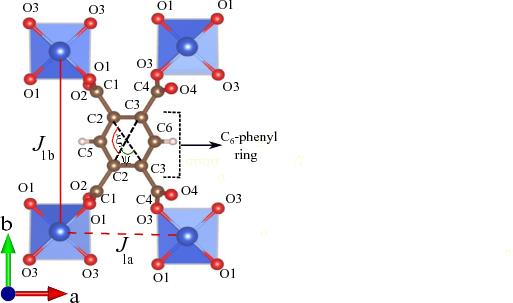}
	\caption \protect{A rectangular unit showing the superexchange interactions $J_{\rm 1a}$ and $J_{\rm 1b}$ along with their respective bridging angles $\angle\psi\simeq60.5^{\circ}$ and $\angle\xi\simeq119.6^{\circ}$ between C-atoms, in the C$_6$-phenyl ring.}
	\label{Fig4}
\end{figure}  
From the crystal structure, the Cu-Cu distance along $b$-direction is greater than the one along $a$-direction. Therefore, one would expect $J_{\rm 1a}$ to be larger than $J_{\rm 1b}$. Similar scenario has been realized in Cu[C$_6$H$_2$(COO)$_4$][C$_2$H$_5$NH$_3$]$_2$ in which the DFT calculations show that $J_{\rm 1a} < J_{\rm 1c}$, even though the Cu-Cu distance along $a$-direction is alomost half of the distance along $c$-direction.\cite{Nath054409} This non-trivial behaviour is attributed to the characteristic features of [C$_6$H$_2$(COO)$_4$]$^{4-}$ anion through which the superexchange takes place. In Cu[C$_6$H$_2$(COO)$_4$][C$_2$H$_5$NH$_3$]$_2$, the effective bridging angles between C atoms belonging to the C$_6$-phenyl ring along the superexchange paths are $\angle\psi\simeq59.9^{\circ}$ and $\angle\xi\simeq120.1^{\circ}$ for $J_{\rm 1a}$ and $J_{\rm 1c}$, respectively in the $ac$-plane. Therefore, it is argued that according to Goodenough-Kanamori-Anderson rules one finds $J_{\rm 1c}>J_{\rm 1a}$ and does not follow Cu-Cu distance. As shown in Fig.~\ref{Fig4}, in C$_{12}$H$_{18}$CuN$_{2}$O$_{11}$, the angles are $\angle\psi\simeq60.5^{\circ}$ and $\angle\xi\simeq119.6^{\circ}$. This explains why $J_{\rm 1a}$ and $J_{\rm 1b}$ have nearly equal values despite different Cu-Cu distances. However, to establish this proposition, a precise estimation of exchange couplings using band structure calculation is required.

%For a nonfrustrated 2D-AF square lattice, the maximum value of the $\chi$ at around the temperature $T_{\chi}^{\rm max}=0.94J_{\rm 2D}$. For this compound the expected value of $T_{\chi}^{\rm max}=5.03$~K which, is close to the experimental $T_{\chi}^{\rm max}=5.1$~K.

%\begin{figure}
%	\includegraphics[width=\linewidth]{Fig6.eps}
%	\caption {Classical phase diagram of $J{_1a,b}-J_2$ model as a function of the frustration angle $\phi$ and anisotropy angle $\theta$. The + sign indicates the position of C$_{12}$H$_{18}$CuN$_{2}$O$_{11}$ (red) and CuPM(EA)$_2$ (black) compounds in this phase diagram}
%	\label{Fig6}
%\end{figure}   

In summary, we have synthesized single crystals of C$_{12}$H$_{18}$CuN$_{2}$O$_{11}$ and reported its crystal structure and magnetic properties in detail. C$_{12}$H$_{18}$CuN$_{2}$O$_{11}$ crystallizes in a monoclinic crystal structure with space group $C2/m$. Because of the low symmetry crystal structure,  Cu$^{2+}$ ions form anisotropic square lattices. The analysis of $\chi(T)$ demonstrates that the compound behaves as a nearly nonfrustrated spin-$1/2$ square lattice with $J_1/k_{\rm B} \simeq 5.3$~K, despite its anisotropic (or rectangular) structural arrangement. Further, the shape of the magnetic isotherm at $T = 1.9$~K could be reproduced well by the QMC simulation assuming a non-frustrated square lattice with $J/k_{\rm B} = 5.2$~K, supporting the $\chi(T)$ analysis. No sign of magnetic LRO down to 2~K indicates minuscule inter-plane coupling in the system. In this compound $J_2$ is negligibly small, but its strength can be increased and the frustration ratio $J_2/J_1$ can be tuned by an appropriate choice of the organic ligand that provides superexchange pathway between the magnetic ions. Thus, the metal organic complexes can reciprocate the inorganic compounds as the model systems in the $J_1 - J_2$ phase diagram.

\section{Acknowledgement}
SG and RN acknowledge SERB, India for financial support bearing sanction order no.~CRG/2019/000960. We thank Alex Andrews for his help in solving the crystal structure.

%\bibliography{reff}


\end{document}